\documentclass[conference]{IEEEtran}
\IEEEoverridecommandlockouts
\usepackage{cite}
\usepackage{amsmath,amssymb,amsfonts}
\usepackage{algorithmic}
\usepackage{graphicx}
\usepackage{textcomp}
\usepackage{xcolor}
\usepackage{url}
\def\BibTeX{{\rm B\kern-.05em{\sc i\kern-.025em b}\kern-.08em
    T\kern-.1667em\lower.7ex\hbox{E}\kern-.125emX}}
\begin{document}

\title{Abductive Corroboration of Probabilistic AI Models for Forensic Synthetic Media Detection
\thanks{This research was funded by NCSC, part of GCHQ.}
}

\author{\IEEEauthorblockN{Junade Ali}
\IEEEauthorblockA{\textit{The Alan Turing Institute} \\
\textit{Defence and National Security Programme}\\
London, United Kingdom \\
jali@turing.ac.uk}
}

\maketitle

\begin{abstract}
Artificial Intelligence (AI) models, at their core, apply general learnings from broad datasets to individual circumstances using probabilistic behaviour. This inductive approach stands in contrast to deductive reasoning approaches which seek to prove conclusions from their premises. However, research has shown that deductive reasoning with AI models is a challenging problem and in the real-world it may not always be feasible. An alternative way forward is to leverage abductive reasoning, seeking to corroborate the output of multiple approaches to identify the most likely conclusion from the factual matrix. We apply this to synthetic media detection in forensic settings, and find we are able to disproportionately lower the risk of false positives to true positive recall. We also provide the first empirical evaluation of OpenAI's rollout of SynthID on synthetic images and evaluate how complementary different synthetic media detection approaches are.
\end{abstract}


\section{Introduction}
Synthetic media presents a challenge for digital forensics practitioners in legal proceedings~\cite{delfino2023deepfakes, chesney2019deep}. Probabilistic models have emerged to detect synthetic media~\cite{yan2023deepfakebench, park2025community, ren2026well, ai-image-detector-2025, Ateeqq, ha2024organic, guo2026ai}, however such models often treat the risk of a false positive as equivalent to the gain of a true positive. The legal burden-of-proof and real-world risk/reward calculation can, however, differ~\cite{chesney2019deep, ali2026r}. It is unknown as to whether such detection models are sufficiently independent in detecting synthetic media from other models and whether cross-corroborating the outputs of different models is sufficient to disproportionately lower false-positive to true-positive recall. Additionally, OpenAI have recently announced SynthID watermarking in synthetic images~\cite{OpenAI, dathathri2024scalable, gowal2025synthid}, however it is not known as to when they began watermarking such images in practice. We therefore seek to answer the following research questions in this paper:

\begin{itemize}
    \item \textbf{RQ1:} When did OpenAI begin adding SynthID to GPT-Image-2 generated images?
    \item \textbf{RQ2:} Across different synthetic media detection approaches, how much similarity is there in the images that are detected?
    \item \textbf{RQ3:} How does cross-corroborating the output of synthetic media detection classifiers affect the ratio between false positive detections to true positive detections?
\end{itemize}

\section{Related Work}
\subsection{AI Reasoning Approaches}
At their core, Artificial Intelligence (AI) technologies are rooted in inductive reasoning~\cite{Mitchell:1980}. In essence, they are trained with a corpus of data, and apply general learnings to specific situations~\cite{lecun2015deep}.

Deductive reasoning, generating a provably correct answer from general rules, has long been a problem in computer science. In 1936 Turing~\cite{turing1936computable} demonstrated a computer cannot be a flawless, terminating solver of every deductive problem.

Experimentally, in the era of Large Language Models (LLMs), Apple's ``The Illusion of Thinking'' paper~\cite{shojaee2026illusion} demonstrated failure cases in high-reasoning tasks amongst LLMs. Natural language reasoning models have improved dramatically, but remain below the human ceiling~\cite{chen2025justlogic}. Additionally, recent work~\cite{cox2026decoding} has highlighted that chain-of-thought reasoning models will often choose an answer first and then use reasoning to generate a chain-of-thought argument in support of it.

However, an alternative approach to lowering AI failure rates exists. Pikies \& Ali at Cloudflare~\cite{pikies2021analysis} demonstrated in 2021 that by placing a second-line good/bad Convolutional Neural Network (CNN) classifier behind traditional string similarity algorithms, false positive rates could be decreased asymmetrically to the cost on true positive performance. In other words, whilst validation leads to a failure to act upon some true positive results, it disproportionately mitigates false positives. This is particularly of use when some high-risk decision is made as a result of some AI output.

This can be considered a form of abductive reasoning, i.e. from a multi-signal factual matrix identifying the simplest and most likely explanation. This has been applied to LLMs under various scenarios. For example, ReConcile~\cite{chen2024reconcile} uses multiple LLMs to form an initial opinion, and then the LLMs are provided visibility of each other's answers and engage in debate to arrive at a final consensus answer. CrossCheckGPT~\cite{sun2024crosscheckgpt} evaluates low consistency between models to identify potential sources of hallucination. FUSE~\cite{lee2026fuse} leverages multiple candidate answers and multiple verifiers to estimate verifier reliability and then identify the best supported answer.

\subsection{Deep Fakes in Forensic Evidence}
Synthetic media poses a challenge for legal decision making. One challenge is introducing fabricated evidence into proceedings~\cite{delfino2023deepfakes}. Another challenge is described by Chesney \& Citron~\cite{chesney2019deep} as the ``liar's dividend'', where someone confronted with evidence can claim it is AI-generated to create reasonable doubt or move the burden-of-proof to verifying the material is authentic.

The ``liar's dividend'' represents an extension of the ``Trojan Horse Defence'' or the ``SODDI Defence'' (the ``Some Other Dude Did It Defence''). Whilst approaches have been developed to address this in digital forensics in child sexual abuse material cases~\cite{ali2026r}, there is limited work considering this in the emerging area of allegations of deep-fake evidence. This can be especially pertinent in cases where the provenance of media is not known due to a lack of verifiable chain-of-custody.

Whilst so far we've discussed reasoning approaches applicable in AI development, rules on determining the admissibility and reliability of evidence can vary by jurisdiction. For example, as outlined in Raitt on Evidence~\cite{raitt-evidence-2018}: ``Scotland is unique in retaining a general principle that crucial facts in criminal cases cannot be proved by the testimony of one witness alone, but must be corroborated by evidence derived from a second independent source.''

Whilst the accuracy score of AI models is often computed with equal weight to true positive, false positive, true negative and false negative results, the real-world risks of a mistaken decision on the basis of such a result can vary. Professional engineering guidance often focuses on judgement in balancing the competing forces of risk and reward. In the Guidance on Risk~\cite{engc2021risk} for the UK engineering professions regulator, the Engineering Council, it is stated: ``Therefore it is important for engineering professionals to understand the level of risk that is acceptable in pursuit of objectives – the risk appetite. Risk appetite defines the boundaries within which risk-based decision making can occur, be controlled and expectations set. Some elements of an engineering activity will be non-negotiable and there will be no appetite for risk. In all cases, there is a need for engineering professionals to exercise informed judgment and leadership in order to manage the risk, consistent with their organisation’s defined risk appetite. All risk decisions should be pursued in line with risk appetite.''

\subsection{Deep-Fake Detection}
Multiple approaches exist for deep-fake detection. The leading approach has so far been to use detector models to evaluate whether media is synthetic or human. DeepFakeBench initially focussed on benchmarking models detecting facial forgery~\cite{yan2023deepfakebench}, however general purpose AI-media detectors have started to emerge - accordingly, many such general purpose models lack independent benchmarking, particularly in out-of-domain experiments.

Independent benchmarks can differ from those originally published. For example, Community Forensics~\cite{park2025community} originally claimed an accuracy of 89.2\% but later independent benchmarking~\cite{ren2026well} reported an accuracy of 75\% (though one table in the pre-print describes the accuracy as 78\% instead). However, there exist other models~\cite{ai-image-detector-2025, Ateeqq} for which we were unable to identify independent benchmarking.

Closed-source vendor models have also been subject to evaluation. For example, a model by Hive Moderation has been subject to analysis~\cite{ha2024organic} which focussed on artwork (280 human-generated and 350 AI-generated) producing an overall accuracy of 98.03\% and later~\cite{guo2026ai} of a sample of 200 images (100 human-generated and 100 AI-generated) an accuracy of 100\%.

An alternative approach has been C2PA metadata~\cite{rosenthol2022c2pa}, where metadata is applied to an AI-generated media file. However, this can trivially be removed either intentionally or unintentionally (e.g. uploading to a website which strips such data~\cite{zewde2026gptimage2}).

Google DeepMind have also produced an approach known as SynthID~\cite{dathathri2024scalable} for watermarking synthetic media. Such watermarking is dependent on the generator applying this watermark. On the 19th May 2026, OpenAI~\cite{OpenAI} reported that they were including such watermarks in images generated through their models, but the exact rollout date is unclear. Whilst Google~\cite{gowal2025synthid} have reported self-evaluated accuracy of a derivative partner deployment of SynthID (SynthID-O) at a configured fixed 0.1\% false positive rate, there is no evaluation of OpenAI's production deployment of SynthID.

\subsection{Literature Gap}
We have identified three key gaps in synthetic media detection. Firstly, there has been little independent investigation of SynthID and its accuracy. One key open question in relation to OpenAI's deployment is when images began being watermarked.

Secondly, there are models which remain without independent benchmarking in out-of-domain contexts. We further have not identified any literature evaluating the extent to which probabilistic detection models form independent judgements.

Finally, we have not identified any literature on whether corroboration of multi-model outputs can reduce the risk of false positives and what the impact is on detection rates generally. Answering such questions can aid in the forensic identification of synthetic media, assisting in the development of reasoning in other areas of AI development.

\section{Methodology}
The following synthetic media detection approaches are used in this paper:

\begin{itemize}
    \item \textbf{M-A}: Fine-tuned SigLIP2 classifier (60,000 AI-generated images and 60,000 human-generated) \cite{Ateeqq}.
    \item \textbf{M-B}: SigLIP2 \& DINOv2 Ensemble model (OpenFake training data, $\approx 95,000$ images for training) \cite{ai-image-detector-2025}.
    \item \textbf{M-C}: Community Forensics \textit{commfor-model-384} classification model (2.7 million images, including 4803 generators)~\cite{park2025community}.
    \item \textbf{M-H}: Hive Moderation API~\cite{ha2024organic, guo2026ai}.
    \item \textbf{SynthID}: OpenAI deployment of SynthID~\cite{OpenAI}.
\end{itemize}

The following data sources were used for evaluation data:

\begin{itemize}
    \item 2000 GPT-Image-2 AI-generated images sourced from X (formerly Twitter) 21st April – 28th April 2026 (no cryptographically verified provenance due to C2PA metadata stripping)~\cite{zewde2026gptimage2}.
    \item 2000 human-generated images (Flickr material uploaded between 2004 and early 2014)~\cite{thomee2016yfcc100m}
\end{itemize}

As the local models (M-\{A,B,C\}) were trained prior to the launch of GPT-Image-2 on the 21st April 2026~\cite{OpenAI_Images2}, GPT-Image-2 generated media should be out-of-domain for all; however, for M-H, secondary classification results did indicate the model was in-distribution.

From this point, all 4000 were classified using the local models (M-\{A,B,C\}). Due to Hive Moderation limiting API queries to 100/day, over $\approx 4$ days, 400 images were classified (200 AI-generated, 200 non-AI). Secondary classification scores (i.e. which particular model was used) were also stored. As OpenAI do not offer a SynthID API and we identified restrictive rate limiting, 400 images were graded over the course of $\approx 4$ days using a headless Chrome browser (the same subset as those subject to the Hive Moderation API).

We then investigated as to when there was a date at which images began being positively classified as containing the SynthID watermark, the independence of synthetic media detector models and the performance of abductive approaches.

\section{RQ1: OpenAI SynthID Investigation}

None of the non-AI images were detected as containing a SynthID watermark. 36 of those which were AI generated were disclosed as containing a SynthID watermark by OpenAI (see Table \ref{tab:synthid_by_source}). Note that C2PA metadata results were all negative, as metadata is removed when a media file is uploaded to X~\cite{zewde2026gptimage2}.

\begin{table}[h]
\centering
\renewcommand{\arraystretch}{1.15}
\caption{OpenAI provenance detection by data source.}
\label{tab:synthid_by_source}
\begin{tabular}{|l|r|r|r|}
\hline
\textbf{Data source} & \textbf{N} & \textbf{SynthID} & \textbf{SynthID \%} \\ \hline
GPT-Image-2 (AI) & 200 & 36 & 18.0 \\ \hline
Multimedia Commons (real) & 200 & 0 & 0.0 \\ \hline
\end{tabular}
\\[3pt]
{\footnotesize OpenAI-covered images only (N=400).}
\end{table}

\begin{figure}[h]
  \centering
  \includegraphics{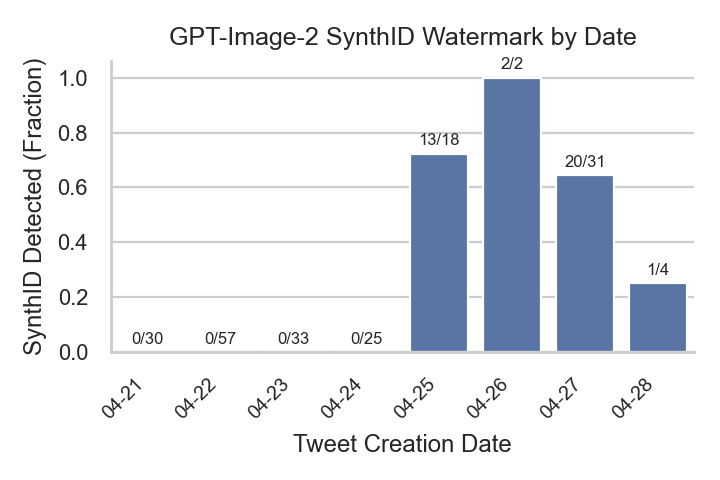}
  \caption{AI-generated Images w/ SynthID by Tweet Post Date}
  \label{fig:synth_id_post_date}
\end{figure}

The dataset provides a date as to when the image was shared on X (formerly Twitter), indicating that images containing SynthID began to appear on the 25th April 2026 (as shown in Table \ref{tab:synthid_by_date} and Figure \ref{fig:synth_id_post_date}). This is just under a month before OpenAI publicly announced this on the 19th May 2026~\cite{OpenAI}.

\begin{table}[h]
\centering
\renewcommand{\arraystretch}{1.15}
\caption{SynthID detection rate by tweet creation date.}
\label{tab:synthid_by_date}
\begin{tabular}{|l|r|r|r|}
\hline
\textbf{Tweet date} & \textbf{N (AI)} & \textbf{SynthID} & \textbf{\%} \\ \hline
2026-04-21 & 30 & 0 & 0.0 \\ \hline
2026-04-22 & 57 & 0 & 0.0 \\ \hline
2026-04-23 & 33 & 0 & 0.0 \\ \hline
2026-04-24 & 25 & 0 & 0.0 \\ \hline
2026-04-25 & 18 & 13 & 72.2 \\ \hline
2026-04-26 & 2 & 2 & 100.0 \\ \hline
2026-04-27 & 31 & 20 & 64.5 \\ \hline
2026-04-28 & 4 & 1 & 25.0 \\ \hline
\end{tabular}
\\[3pt]
{\footnotesize GPT-Image-2 images with OpenAI coverage and a known date (N=200).}
\end{table}

\section{RQ2: Independence of Synthetic Media Detectors}
Across 4000 images, we saw varying performance from the three open-source models, as shown in Table \ref{tab:metrics_3full}.

\begin{table}[h]
\centering
\renewcommand{\arraystretch}{1.15}
\caption{Metrics for open-source detectors.}
\label{tab:metrics_3full}
\begin{tabular}{|l|r|r|r|r|r|r|}
\hline
\textbf{Detector} & \textbf{TP} & \textbf{FP} & \textbf{Recall \%} & \textbf{FPR \%} & \textbf{Prec. \%} & \textbf{Acc. \%} \\ \hline
M-C & 150 & 1 & 7.5 & 0.1 & 99.3 & 53.7 \\ \hline
M-A & 1701 & 336 & 85.0 & 16.8 & 83.5 & 84.1 \\ \hline
M-B & 1582 & 110 & 79.1 & 5.5 & 93.5 & \textbf{86.8} \\ \hline
\end{tabular}
\\[3pt]
{\footnotesize N=4000 (2000 AI / 2000 real).}
\end{table}

\begin{figure}[h]
  \centering
  \includegraphics{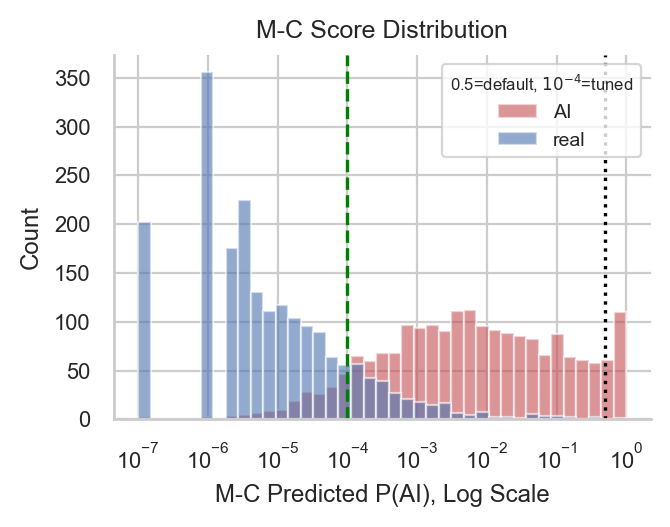}
  \caption{M-C Score Distribution}
  \label{fig:fig_calibration_dist}
\end{figure}

\begin{figure}[h]
  \centering
  \includegraphics{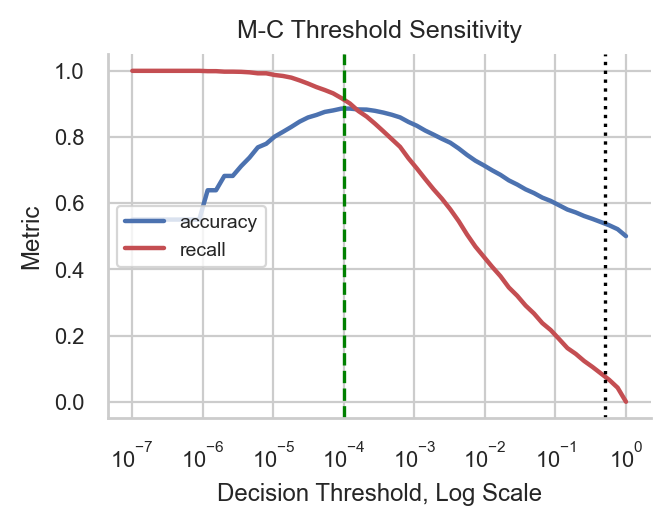}
  \caption{M-C Threshold Sensitivity}
  \label{fig:fig_calibration_thresh}
\end{figure}

\begin{figure}[h]
  \centering
  \includegraphics{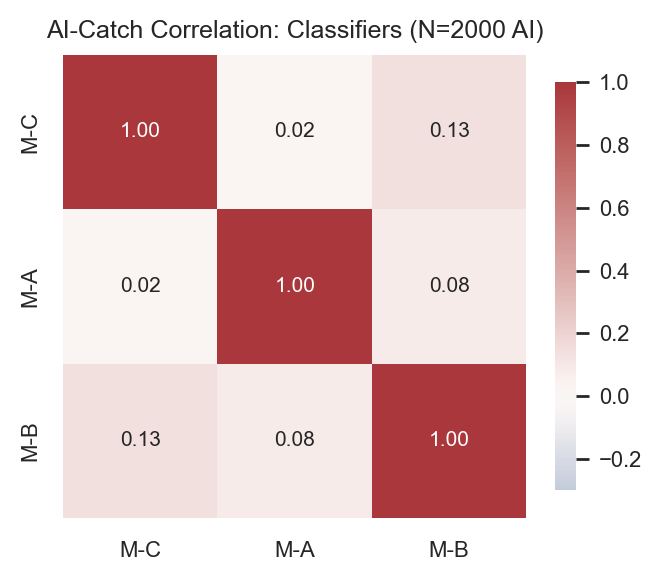}
  \caption{$\phi$ correlation of recall on AI images (open-source models)}
  \label{fig:fig_independence_3}
\end{figure}

Note that the M-C model saw an exceptionally low false positive rate, but lower true positive count than any other model. We identified that the reason for this was the model was poorly calibrated on this out-of-domain dataset. As shown in Figure \ref{fig:fig_calibration_dist}, there is a separation between the AI and non-AI images, but at a very low threshold. As shown in Figure \ref{fig:fig_calibration_thresh}, it is possible to adjust this threshold for optimisation purposes on this model. $\phi$ correlation of recall on AI images is shown in Figure \ref{fig:fig_independence_3}, a $\phi$ of 1 indicates that the detected AI images are effectively the same as another detector whilst a $\phi$ closest to 0 indicates the models are fully complementary.

\begin{figure}[h]
  \centering
  \includegraphics{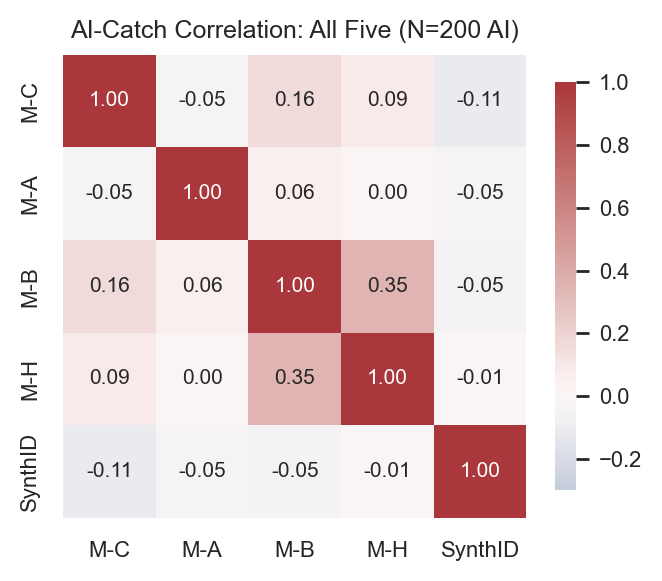}
  \caption{$\phi$ correlation of recall on AI images (all detectors)}
  \label{fig:fig_independence_5}
\end{figure}

Turning next to the subset of 400 images which were subject to the open-source models, Hive Moderation (M-H) and the SynthID verification, both M-C and M-H saw 0\% false positive rates on this dataset as shown in Table \ref{tab:metrics_prob4}. As we saw across the entire dataset, M-C did yield a false positive, despite a very low recall rate, so accordingly this should not be taken to assume that M-H has an absolute zero false positive rate across larger datasets - however it does indicate it is very low (despite strong recall). However, as shown in Table \ref{tab:detection_by_model} we found that model M-A saw the most cases where it was the only model to correctly positively identify an AI-generated image. The $\phi$ correlation of recall on AI images is shown in Figure \ref{fig:fig_independence_5}, including SynthID. The negative correlations on SynthID detectors are expected as a majority of the AI-generated images did not have the SynthID watermark embedded during OpenAI's rollout.

\begin{table}[h]
\centering
\renewcommand{\arraystretch}{1.15}
\caption{Metrics for all probabilistic detectors.}
\label{tab:metrics_prob4}
\begin{tabular}{|l|r|r|r|r|r|r|}
\hline
\textbf{Detector} & \textbf{TP} & \textbf{FP} & \textbf{Recall \%} & \textbf{FPR \%} & \textbf{Prec. \%} & \textbf{Acc. \%} \\ \hline
M-C & 19 & 0 & 9.5 & 0.0 & 100.0 & 54.8 \\ \hline
M-A & 169 & 38 & 84.5 & 19.0 & 81.6 & 82.8 \\ \hline
M-B & 159 & 22 & 79.5 & 11.0 & 87.8 & 84.2 \\ \hline
M-H & 168 & 0 & 84.0 & 0.0 & 100.0 & \textbf{92.0} \\ \hline
\end{tabular}
\\[3pt]
{\footnotesize N=400 (200 AI / 200 real).}
\end{table}

\begin{table}[h]
\centering
\renewcommand{\arraystretch}{1.15}
\caption{Count where detector was the \emph{only} true positive.}
\label{tab:detection_by_model}
\begin{tabular}{|l|r|r|r|}
\hline
\textbf{Detector} & \textbf{Caught} & \textbf{Recall \%} & \textbf{Sole detector} \\ \hline
M-C & 19 & 9.5 & 0 \\ \hline
M-A & 169 & 84.5 & \textbf{10} \\ \hline
M-B & 159 & 79.5 & 0 \\ \hline
M-H & 168 & 84.0 & 3 \\ \hline
SynthID & 36 & 18.0 & 0 \\ \hline
\end{tabular}
\\[3pt]
{\footnotesize Common subset, N=200 AI images.}
\end{table}

Finally, Table \ref{tab:hive_generators} shows the top generator attribution from Hive Moderation of images it attributed to be AI-generated. The fact that the top attributed model is \textit{gptimage2} indicates that GPT-Image-2 was in-distribution for the model.

\begin{table}[h]
\centering
\renewcommand{\arraystretch}{1.15}
\caption{Hive (M-H) top source-generator attribution.}
\label{tab:hive_generators}
\begin{tabular}{|l|r|r|r|}
\hline
\textbf{Hive generator class} & \textbf{Top-1 count} & \textbf{\% of det.} & \textbf{Mean score} \\ \hline
gptimage2 & 117 & 69.6 & 0.918 \\ \hline
gptimage1\_5 & 12 & 7.1 & 0.633 \\ \hline
reve & 9 & 5.4 & 0.692 \\ \hline
gemini3 & 6 & 3.6 & 0.830 \\ \hline
4o & 5 & 3.0 & 0.584 \\ \hline
stablediffusion & 5 & 3.0 & 0.527 \\ \hline
flux & 4 & 2.4 & 0.573 \\ \hline
qwen & 2 & 1.2 & 0.720 \\ \hline
\end{tabular}
\\[3pt]
{\footnotesize Hive-detected GPT-Image-2 images (N=168).}
\end{table}

\section{RQ3: Corroborated Approaches}
On the subset of 400 images, we saw a rapid decrease in false positive rate from 28\% to 2\% simply by relying on the output of two detectors instead of just one. This dropped to 0\% when corroboration spanned three detectors. This is shown in Table \ref{tab:corrob_all5}. Interestingly, if the most disparate probabilistic detectors alone required corroboration (M-C and M-A), there were no false positives, but the recall rate dropped to just 7.5\% as only 15 true positives were recorded. The absolute accuracy rate where any model was corroborated with the output of any other was 94.8\%, beating even M-H at 92.0\% accuracy.

\begin{table}[h]
\centering
\renewcommand{\arraystretch}{1.15}
\caption{Corroboration over all five detectors.}
\label{tab:corrob_all5}
\begin{tabular}{|l|r|r|r|r|r|r|}
\hline
\textbf{Rule} & \textbf{TP} & \textbf{FP} & \textbf{Recall \%} & \textbf{FPR \%} & \textbf{Prec. \%} & \textbf{Acc. \%} \\ \hline
$\geq$1 & 196 & 56 & 98.0 & 28.0 & 77.8 & 85.0 \\ \hline
$\geq$2 & 183 & 4 & 91.5 & 2.0 & 97.9 & \textbf{94.8} \\ \hline
$\geq$3 & 138 & 0 & 69.0 & 0.0 & 100.0 & 84.5 \\ \hline
$\geq$4 & 34 & 0 & 17.0 & 0.0 & 100.0 & 58.5 \\ \hline
$\geq$5 & 0 & 0 & 0.0 & 0.0 & -- & 50.0 \\ \hline
\end{tabular}
\\[3pt]
{\footnotesize 5 detectors; N=400 (200 AI / 200 real).}
\end{table}

Where the out-of-domain detectors were used across all 4000 images, the results are shown in Table \ref{tab:corrob_3}. In terms of accuracy metric, an uncorroborated finding from any one model beat the performance of any individual model, however note that the false-positive rate under this configuration was significant (20.7\%). By contrast, requiring corroboration between two models dropped this false positive rate to 1.7\% and to 0\% between three models.

\begin{table}[h]
\centering
\renewcommand{\arraystretch}{1.15}
\caption{Corroboration over the three classifiers (full dataset).}
\label{tab:corrob_3}
\begin{tabular}{|l|r|r|r|r|r|r|}
\hline
\textbf{Rule} & \textbf{TP} & \textbf{FP} & \textbf{Recall \%} & \textbf{FPR \%} & \textbf{Prec. \%} & \textbf{Acc. \%} \\ \hline
$\geq$1 & 1914 & 414 & 95.7 & 20.7 & 82.2 & \textbf{87.5} \\ \hline
$\geq$2 & 1389 & 33 & 69.5 & 1.7 & 97.7 & 83.9 \\ \hline
$\geq$3 & 130 & 0 & 6.5 & 0.0 & 100.0 & 53.2 \\ \hline
\end{tabular}
\\[3pt]
{\footnotesize 3 detectors; N=4000 (2000 AI / 2000 real).}
\end{table}

This effect can be seen in Table \ref{tab:eff_corrob_prob4} (for all four detectors) and Table \ref{tab:eff_corrob_3} (for the three out-of-domain detectors), where a FP/TP rate is calculated (i.e. the cost of obtaining a true positive for each false positive, a higher value indicates a greater cost).

\begin{table}[h]
\centering
\renewcommand{\arraystretch}{1.15}
\caption{FP/TP over all four probabilistic detectors.}
\label{tab:eff_corrob_prob4}
\begin{tabular}{|l|r|r|r|}
\hline
\textbf{Rule} & \textbf{TP} & \textbf{FP} & \textbf{FP/TP} \\ \hline
$\geq$1 & 196 & 56 & 0.29 \\ \hline
$\geq$2 & 178 & 4 & 0.02 \\ \hline
$\geq$3 & 127 & 0 & 0.00 \\ \hline
$\geq$4 & 14 & 0 & 0.00 \\ \hline
M-C $\vee$ M-A & 173 & 38 & 0.22 \\ \hline
M-C $\wedge$ M-A & 15 & 0 & 0.00 \\ \hline
\end{tabular}
\\[3pt]
{\footnotesize 4 detectors; N=400. FP/TP = false positives per true positive (0.00 = zero FP). Final two rows: most-disparate pair M-C, M-A.}
\end{table}

\begin{table}[h]
\centering
\renewcommand{\arraystretch}{1.15}
\caption{FP/TP over the out-of-domain probabilistic classifiers.}
\label{tab:eff_corrob_3}
\begin{tabular}{|l|r|r|r|}
\hline
\textbf{Rule} & \textbf{TP} & \textbf{FP} & \textbf{FP/TP} \\ \hline
$\geq$1 & 1914 & 414 & 0.22 \\ \hline
$\geq$2 & 1389 & 33 & 0.02 \\ \hline
$\geq$3 & 130 & 0 & 0.00 \\ \hline
M-C $\vee$ M-A & 1719 & 337 & 0.20 \\ \hline
M-C $\wedge$ M-A & 132 & 0 & 0.00 \\ \hline
\end{tabular}
\\[3pt]
{\footnotesize 3 detectors; N=4000. FP/TP = false positives per true positive (0.00 = zero FP). Final two rows: most-disparate pair M-C, M-A.}
\end{table}

\section{Analysis \& Limitations}
In relation to \textbf{RQ1}, in the week following the launch of GPT-Image-2 by OpenAI, we found evidence that SynthID watermarks began appearing on the 25th April 2026. Of 200 AI-generated images, 36 contained this watermark. Of 200 non-AI images, OpenAI found none of these images contained the watermark. This finding is useful for forensic practitioners in demonstrating that even before official launch, SynthID watermarking can be useful to assess the provenance of such images (but should not be considered conclusive).

In relation to \textbf{RQ2}, across the five detectors evaluated, M-H (Hive Moderation) and M-B showed the greatest similarity in images they would detect, however still only yielded a $\phi$ correlation of 0.35. For probabilistic detectors, M-A and M-C yielded the greatest dissimilarity. The probabilistic detectors all yielded varying performance when considering accuracy and precision.  Future work may wish to consider when detectors are running in out-of-domain circumstances, how recalibration could affect these metrics.

In relation to \textbf{RQ3}, across the three detectors on 4000 images, we found that requiring corroboration between the output of two detectors reduced the ratio of false positive to true positive detections (FP/TP) from 0.22 to 0.02. Increasing the corroboration to three models reduced this ratio to 0.00. A similar effect was seen in the subset of 400 images classified using five detectors.

Future work may wish to consider this on alternative datasets, such as an alternative newer dataset of human-generated images and other synthetic media generators. We note that a single model operated at a stricter threshold also trades recall for precision; a comparison of corroboration against threshold-tuned single models and score-level fusion is left to future work. Where rates of zero are reported, the 95\% Clopper–Pearson upper bound is 1.83\% for N=200 and 0.18\% for N=2000.

\section{Conclusion}
This paper highlights the importance of diversity of reasoning in Artificial Intelligence models. Whilst there has historically been a tendency to consider that larger singular models are better, our research here has shown that corroboration of the output of multiple models can produce disproportionately lower false positives to true positives, even in the context of synthetic media detection.

For forensic practitioners specifically, this provides quantifiable results showing this phenomenon in this context. This paper also provides forensic practitioners, for the first time, quantifiable results showing OpenAI's SynthID watermark demonstrated no false positives for synthetic media detection and evidence that such watermarking was in operation before the official launch.

In relation to Artificial Intelligence development, accuracy metrics have typically considered false positives and true positives to be of equal weight. Where risk exists in the case of an erroneous output or a legal burden-of-proof is different, this metric may not be applicable. This paper highlights that there exists a route beyond this inductive reasoning approach, even where deductive conclusions cannot be reached. Abductive approaches, which seek to identify the most likely conclusion based on corroborating the output of multiple models, provide a path forward.

However, a key limitation (and area for future work) is that in a world where models become larger and there is greater potential for overlap in training data, it is unclear whether such approaches require the creation of more models in more domains (such as LLMs) that share less similarity in design and training.

\bibliographystyle{plain} 
\bibliography{bibliography}

@InCollection{Mitchell:1980,
  author =       "Mitchell, Tom M.",
  title =        "The Need for Biases in Learning Generalizations",
  booktitle =    "Readings in Machine Learning",
  publisher =    "Morgan Kauffman",
  year =         "1980",
  editor =    "Shavlik, Jude W. and Dietterich, Thomas G.",
  pages =     "184--191",
  url = "https://www.cs.cmu.edu/afs/cs/usr/mitchell/ftp/pubs/NeedForBias_1980.pdf",
  bib2html_rescat = "General ML",
  note = {\url{https://www.cs.cmu.edu/afs/cs/usr/mitchell/ftp/pubs/NeedForBias_1980.pdf} Accessed: 2026-06-27}
}

@article{lecun2015deep,
  title={Deep learning},
  author={LeCun, Yann and Bengio, Yoshua and Hinton, Geoffrey},
  journal={nature},
  volume={521},
  number={7553},
  pages={436--444},
  year={2015},
  publisher={Nature Publishing Group UK London}
}

@article{shojaee2026illusion,
  title={The illusion of thinking: Understanding the strengths and limitations of reasoning models via the lens of problem complexity},
  author={Shojaee, Parshin and Mirzadeh, Iman and Horton, Maxwell and Bengio, Samy and Farajtabar, Mehrdad and others},
  journal={Advances in Neural Information Processing Systems},
  volume={38},
  pages={108018--108059},
  year={2026}
}

@article{chen2025justlogic,
  title={Justlogic: A comprehensive benchmark for evaluating deductive reasoning in large language models},
  author={Chen, Michael K and Zhang, Xikun and Tao, Dacheng},
  journal={arXiv preprint arXiv:2501.14851},
  year={2025}
}

@article{cox2026decoding,
  title={Decoding Answers Before Chain-of-Thought: Evidence from Pre-CoT Probes and Activation Steering},
  author={Cox, Kyle and Kianersi, Darius and Garriga-Alonso, Adri{\`a}},
  journal={arXiv preprint arXiv:2603.01437},
  year={2026}
}

@article{turing1936computable,
  title={On computable numbers, with an application to the Entscheidungsproblem},
  author={Turing, Alan Mathison and others},
  journal={J. of Math},
  volume={58},
  number={345-363},
  pages={5},
  year={1936},
  publisher={Wiley Online Library}
}

@article{pikies2021analysis,
  title={Analysis and safety engineering of fuzzy string matching algorithms},
  author={Pikies, Malgorzata and Ali, Junade},
  journal={ISA transactions},
  volume={113},
  pages={1--8},
  year={2021},
  publisher={Elsevier}
}

@inproceedings{chen2024reconcile,
  title={Reconcile: Round-table conference improves reasoning via consensus among diverse llms},
  author={Chen, Justin and Saha, Swarnadeep and Bansal, Mohit},
  booktitle={Proceedings of the 62nd Annual Meeting of the Association for Computational Linguistics (Volume 1: Long Papers)},
  pages={7066--7085},
  year={2024}
}

@article{sun2024crosscheckgpt,
  title={CrossCheckGPT: Universal hallucination ranking for multimodal foundation models},
  author={Sun, Guangzhi and Manakul, Potsawee and Liusie, Adian and Pipatanakul, Kunat and Zhang, Chao and Woodland, Phil and Gales, Mark},
  journal={arXiv preprint arXiv:2405.13684},
  year={2024}
}

@article{lee2026fuse,
  title={FUSE: Ensembling Verifiers with Zero Labeled Data},
  author={Lee, Joonhyuk and Ma, Virginia and Zhao, Sarah and Nair, Yash and Spector, Asher and Cohen, Regev and Cand{\`e}s, Emmanuel J},
  journal={arXiv preprint arXiv:2604.18547},
  year={2026}
}

@book{raitt-evidence-2018,
  title     = {Raitt on Evidence: Principles, Policy and Practice},
  editor    = {Keane, Eamon and Davidson, Fraser},
  edition   = {3rd},
  publisher = {W. Green},
  address   = {Edinburgh},
  year      = {2018},
  isbn      = {9780414032958},
  note      = {Originally authored by Fiona E. Raitt},
}

@article{delfino2023deepfakes,
  title={Deepfakes on trial: a call to expand the trial judge’s gatekeeping role to protect legal proceedings from technological fakery},
  author={Delfino, Rebecca},
  journal={Deepfakes on Trial: A Call To Expand the Trial Judge’s Gatekeeping Role To Protect Legal Proceedings from Technological Fakery},
  volume={74},
  pages={2022--02},
  year={2023}
}

@article{chesney2019deep,
  title={Deep fakes},
  author={Chesney, Bobby and Citron, Danielle},
  journal={California law review},
  volume={107},
  number={6},
  pages={1753--1820},
  year={2019},
  publisher={JSTOR}
}

@inproceedings{ali2026r,
  title={R v F (2025): Addressing the Defence of Hacking},
  author={Ali, Junade},
  booktitle={2026 14th International Symposium on Digital Forensics and Security (ISDFS)},
  pages={1--4},
  year={2026},
  organization={IEEE}
}

@techreport{engc2021risk,
  author      = {{Engineering Council}},
  title       = {Guidance on Risk for the Engineering Profession},
  institution = {Engineering Council},
  year        = {2021},
  month       = oct,
  address     = {London, UK},
  url         = {https://www.engc.org.uk/media/ye2lwps1/guidance-on-risk.pdf},
  urldate     = {2026-06-27},
  note = {\url{https://www.engc.org.uk/media/ye2lwps1/guidance-on-risk.pdf} Accessed: 2026-06-27}
}

@article{yan2023deepfakebench,
  title={Deepfakebench: A comprehensive benchmark of deepfake detection},
  author={Yan, Zhiyuan and Zhang, Yong and Yuan, Xinhang and Lyu, Siwei and Wu, Baoyuan},
  journal={arXiv preprint arXiv:2307.01426},
  year={2023}
}

@inproceedings{park2025community,
  title={Community forensics: Using thousands of generators to train fake image detectors},
  author={Park, Jeongsoo and Owens, Andrew},
  booktitle={Proceedings of the Computer Vision and Pattern Recognition Conference},
  pages={8245--8257},
  year={2025}
}

@article{ren2026well,
  title={How well are open sourced AI-generated image detection models out-of-the-box: A comprehensive benchmark study},
  author={Ren, Simiao and Zhou, Yuchen and Shen, Xingyu and Zewde, Kidus and Duong, Tommy and Huang, George and Wei, En and Xue, Jiayu and others},
  journal={arXiv preprint arXiv:2602.07814},
  year={2026}
}

@misc{ai-image-detector-2025,
  author = {Bombek1},
  title = {AI Image Detector (SigLIP2 + DINOv2 Ensemble)},
  year = {2025},
  publisher = {Hugging Face},
  url = {https://huggingface.co/Bombek1/ai-image-detector-siglip-dinov2},
  note = {\url{https://huggingface.co/Bombek1/ai-image-detector-siglip-dinov2} Accessed: 2026-06-27}
}

@misc{Ateeqq,
  author = {Ateeqq},
  title = {AI and Human Image Classification Model - v1},
  year = {2025},
  publisher = {Hugging Face},
  url = {https://huggingface.co/Ateeqq/ai-vs-human-image-detector},
  note = {\url{https://huggingface.co/Ateeqq/ai-vs-human-image-detector} Accessed: 2026-06-27}
}

@inproceedings{rosenthol2022c2pa,
  title={C2pa: the world’s first industry standard for content provenance (conference presentation)},
  author={Rosenthol, Leonard},
  booktitle={Applications of Digital Image Processing XLV},
  volume={12226},
  pages={122260P},
  year={2022},
  organization={SPIE}
}

@inproceedings{ha2024organic,
  title={Organic or diffused: Can we distinguish human art from ai-generated images?},
  author={Ha, Anna Yoo Jeong and Passananti, Josephine and Bhaskar, Ronik and Shan, Shawn and Southen, Reid and Zheng, Haitao and Zhao, Ben Y},
  booktitle={Proceedings of the 2024 on ACM SIGSAC Conference on Computer and Communications Security},
  pages={4822--4836},
  year={2024}
}

@inproceedings{guo2026ai,
  title={Ai-generated image detection: Passive or watermark?},
  author={Guo, Moyang and Hu, Yuepeng and Jiang, Zhengyuan and Li, Zeyu and Sadovnik, Amir and Daw, Arka and Gong, Neil Zhenqiang},
  booktitle={Proceedings of the IEEE/CVF Conference on Computer Vision and Pattern Recognition},
  pages={400--410},
  year={2026}
}

@misc{OpenAI,
  author = {OpenAI},
  title = {Advancing content provenance for a safer, more transparent AI ecosystem},
  year = {2026},
  publisher = {OpenAI},
  url = {https://openai.com/index/advancing-content-provenance/},
  note = {\url{https://openai.com/index/advancing-content-provenance/} Accessed: 2026-06-27}
}

@article{gowal2025synthid,
  title={SynthID-Image: Image watermarking at internet scale},
  author={Gowal, Sven and Bunel, Rudy and Stimberg, Florian and Stutz, David and Ortiz-Jimenez, Guillermo and Kouridi, Christina and Vecerik, Mel and Hayes, Jamie and Rebuffi, Sylvestre-Alvise and Bernard, Paul and others},
  journal={arXiv preprint arXiv:2510.09263},
  year={2025}
}

@article{dathathri2024scalable,
  title={Scalable watermarking for identifying large language model outputs},
  author={Dathathri, Sumanth and See, Abigail and Ghaisas, Sumedh and Huang, Po-Sen and McAdam, Rob and Welbl, Johannes and Bachani, Vandana and Kaskasoli, Alex and Stanforth, Robert and Matejovicova, Tatiana and others},
  journal={Nature},
  volume={634},
  number={8035},
  pages={818--823},
  year={2024},
  publisher={Nature Publishing Group UK London}
}

@article{zewde2026gptimage2,
  author  = {Zewde, Kidus and Ren, Simiao and Shen, Xingyu and Wu, Jenny
             and Zhou, Yuchen and Duong, Tommy and Zhang, Zikang and Traister, Ethan},
  title   = {{GPT-Image-2} in the Wild: A {Twitter} Dataset of Self-Reported
             {AI}-Generated Images from the First Week of Deployment},
  journal = {arXiv preprint},
  year    = {2026}
}

@article{thomee2016yfcc100m,
  title={Yfcc100m: The new data in multimedia research},
  author={Thomee, Bart and Shamma, David A and Friedland, Gerald and Elizalde, Benjamin and Ni, Karl and Poland, Douglas and Borth, Damian and Li, Li-Jia},
  journal={Communications of the ACM},
  volume={59},
  number={2},
  pages={64--73},
  year={2016},
  publisher={ACM New York, NY, USA}
}

@misc{OpenAI_Images2,
  author = {OpenAI},
  title = {Introducing ChatGPT Images 2.0},
  year = {2026},
  publisher = {OpenAI},
  url = {https://openai.com/index/introducing-chatgpt-images-2-0/},
  note = {\url{https://openai.com/index/introducing-chatgpt-images-2-0/} Accessed: 2026-06-27}
}

\end{document}